\documentclass{IAU}
\usepackage{graphicx}
\usepackage{url}

\title[Proxima Centauri and Trappist 1 planetary systems]
{Migration of bodies in the Proxima Centauri and Trappist 1 planetary systems}

\author[Sergei I. Ipatov]   {Sergei I. Ipatov}

\affiliation{Vernadsky Institute of Geochemistry and Analytical Chemistry of RAS, 
\\ 119991, 19 Kosygin st.,
Moscow, Russia \\ email: {\tt siipatov@hotmail.com}}

\pubyear{2024}
\volume{393}  
\jname{ Planetary Science and Exoplanets in the Era of the James Webb Space Telescope}
\editors{Youssef Moulane, ed.}

\begin{document}

\maketitle

\begin{abstract}
The motion of planetesimals was studied in the Proxima Centauri and TRAPPIST 1 exoplanetary systems. The size of the feeding zone of planet Proxima Centauri c is discussed. It was noted that after hundreds of Myrs, some planetesimals could still move in elliptical resonant orbits inside the feeding zone of this planet that had been mainly cleared from planetesimals. The probability of a collision of a planetesimal initially located in the feeding zone of planet c with inner planet b was obtained to be about 0.0002 and 0.001 at initial eccentricity of orbits of planetesimals equal to 0.02 or 0.15, respectively. A lot of icy material and volatiles could be delivered from the icy zone near the orbit of planet c to inner planets b and d. The inclinations of orbits of 80\% of the planetesimals that moved between 500 or 1200 AU from the star did not exceed $10^o$. It was obtained that several planets in the TRAPPIST-1 system accumulated planetesimals initially located at the same distance. Outer layers of neighbouring TRAPPIST-1 planets can include similar material.  

\end{abstract}

\firstsection 

\section{Initial data for calculations of the motion of planetesimals in the Proxima Centauri planetary system}
The motion of planetesimals and other bodies in the Solar System was studied in many papers (see e.g. the review in (Marov, Ipatov, 2023)). Ipatov (2023a,b,c) studied the motion of planetesimals in the Proxima Centauri planetary system. In these calculations, the gravitational influence of the star (with a mass equal to 0.122 of the solar mass) and two planets: b ($a_b$=0.04857 AU, $e_b$=0.11, $m_b$=1.17$m_E$, $m_E$ is the mass of the Earth) and c ($a_c$=1.489 AU, $e_c$=0.04, $m_c$=7$m_E$) was taken into account. Calculations with other values of $m_c$ (0.7$m_E$, 3.5$m_E$, and 12$m_E$) were also made. The planetesimals were excluded from integration when they collided with the star or the planets or reached 1200 AU (the Hill radius of the star) from the star. The symplectic code from (Levison, Duncan, 1994) was used for integrations. In most calculations, the integration time step $t_s$ equaled to 1 day. The obtained results were about the same for different considered $t_s$ equaled to 0.2 day, or 0.5 day, or 2 days. The considered time interval usually exceeded 100 Myr. For some variants it reached 1000 Myr. The late gas-less stage of formation of planets was considered. Migration of planetesimals from the feeding zone of planet c was calculated. In each calculation variant, initial semi-major axes $a_o$ of orbits of $N_o$=250 planetesimals were in the range from $a_{min}$ to $a_{min}$+0.1 AU. In different variants, the values of $a_{min}$ were in the range from 0.9 to 2.2 AU. Initial eccentricities $e_o$ of orbits of planetesimals equaled to 0.02 or 0.15.  Initial inclinations of orbits of the planetesimals were equal to $e_o$/2 rad. Not small considered eccentricities of planetesimals could be a result of the previous evolution of the disk of planetesimals and could be reached due to mutual gravitational influence of planetesimals.

\section{The feeding zone of Proxima Centauri c}

After hundreds of million years, some planetesimals could still move in elliptical orbits inside the feeding zone of planet c that had been mainly cleared from planetesimals. Initial semi-major axes of orbits of the planetesimals that still had elliptical orbits at the end of considered evolution are presented in \cite[(Ipatov, 2023a)]{Ipatov2023a}. For the range ($a_{mine}$, $a_{maxe}$) of initial semi-major axes $a_o$ of orbits for which planetesimals were mainly ejected into hyperbolic orbits or collided with planets, Ipatov (2023a) obtained $a_{mine}$=$a_c$(1-$e_c$)-$e_o$$a_{mine}$-$k_{min}$$a_c$$\mu^{1/3}$ and $a_{maxe}$=$a_c$(1+$e_c$)+$e_o$$a_{maxe}$+$k_{max}$$a_c$$\mu^{1/3}$, where $a_c$ and $e_c$=0.04 are the semi-major axis and eccentricity of the orbit of planet c, and $\mu$ is the ratio of the mass of planet c to the star mass . In these formulas, $k_{min}$=2.54 and $k_{max}$=2.40 at $e_o$=0.02, and $k_{min}$=2.23 and $k_{max}$=4.3 at $e_o$=0.15.
Some planetesimals could still move inside the feeding zone, often in some resonances with the planet. For example, they moved in the resonances 1:1 (as Jupiter Trojans), 5:4, and 3:4. The number of such left planetesimals was greater at smaller eccentricities. Ipatov (2023a) presented the examples of time evolution of a semi-major axis $a$, perihelion and aphelion distances $q$ and $Q$ of an orbit of a planetesimal moving in such resonances.  Before getting for a long time in these resonances, some planetesimals could move in nonresonant orbits. Some planetesimals could be ejected into hyperbolic orbits or could collide with planets if they started from some (typically resonant) subregions of $a_o$ located outside the main feeding zone of planet c.    

The  ratio of the mass of planet Proxima Centauri c to the mass of the star is smaller than that for Jupiter. The ratio of the semi-major axes of the orbits of the planets c and b is larger than that for Jupiter and Mars. There is only one large planet in the Proxima Centauri system. 
Therefore there  can be more analogues of the asteroid and trans-Neptunian belts in the planetary system near Proxima Centauri than in the Solar System, and there can be  
 a planet (planets) between the orbits of the planets Proxima Centauri b and c.

\section{Probabilities of ejection of planetesimals  into hyperbolic orbits and the total initial mass of planetesimals in the feeding zone of Proxima Centauri c} 

Ipatov (2023b) presented the fraction of initial planetesimals that were left in elliptical orbits after times $T$ equal to 1, 10, and 100 Myr at different initial semi-major axes $a_o$ of their orbits. For $T$=100 Myr and $e_o$=0.15,  
the fraction was not more than 0.3, 0.06, 0.2, 0.06, and 0.1 
at 1.1$\le$$a_o$$\le$1.2 AU, at 1.2$\le$$a_o$$\le$1.7 AU, at 1.7$\le$$a_o$$\le$1.9 AU,  at 2.0$\le$$a_o$$\le$2.1 AU,  and at 2.1$\le$$a_o$$\le$2.2 AU, respectively.  For $e_o$=0.02, this fraction was not more than 0.25 at 1.2$\le$$a_o$$\le$1.8 AU.
If initial orbits of planetesimals were not far from the orbit of the planet c, then the probability of a collision of a planetesimal during its dynamical lifetime with planet c was about 0.5 for $e_o$=0.02 and 0.25-0.3 for $e_o$=0.15. For $e_o$=0.02 the probability was about 0.05, 0.4-0.55, 0.3, and 0.02 at 
1.1$\le$$a_o$$\le$1.2 AU, at 1.2$\le$$a_o$$\le$1.7 AU, at 1.7$\le$$a_o$$\le$1.8 AU, and at 1.8$\le$$a_o$$\le$1.9 AU, respectively. For $e_o$=0.15, 
this probability was about 0.15, 0.3, and 0.04-0.05
at 1.0$\le$$a_o$$\le$1.1 AU, at 1.1$\le$$a_o$$\le$1.9 AU, and at 2.0$\le$$a_o$$\le$2.2 AU, respectively. 
 Most collisions of planetesimals with planet c took place during the first 10 Myr. 

If initial orbits of planetesimals were not far from the orbit of the planet c, then during a whole considered time interval the ratio $p_{cej}$ of the number of planetesimals collided with planet c to that ejected into hyperbolic orbits was about 1 for $e_o$=0.02 and about 0.5 for $e_o$=0.15.
For $e_o$=0.02, the ratio $p_{cej}$ was about 2 at 1.1$\le$$a_o$$\le$1.2 AU, about 1.2 at 1.2$\le$$a_o$$\le$1.4 AU, about 0.85-1 at 1.4$\le$$a_o$$\le$1.6 AU, about 1.3 at 1.6$\le$$a_o$$\le$1.7 AU, about 0.5 at 1.7$\le$$a_o$$\le$1.8 AU, and about 0.2 at 1.8$\le$$a_o$$\le$1.9 AU. For $e_o$=0.15, this ratio was about 0.9 at 1.0$\le$$a_o$$\le$1.2 AU, 0.5-0.6 at 1.2$\le$$a_o$$\le$1.4 AU, 0.4 at 1.4$\le$$a_o$$\le$1.8 AU, 0.05 at 2.0$\le$$a_o$$\le$2.2 AU, and 0.02 at 2.2$\le$$a_o$$\le$2.3 AU. The ratio of the fraction of planetesimals ejected into hyperbolic orbits to that collided with exoplanets is greater for a greater mass of a planet moving in the orbit of planet c and for greater initial eccentricities of orbits of planetesimals. 

For a whole feeding zone of planet c and the present mass of planet c (7$m_E$), Ipatov (2023b) concluded that the ratio $p_{cej}$=$p_c/p_{ej}$ of the probability of a collision of a 
planetesimal with planet c to the probability $p_{ej}$ of its ejection into a hyperbolic orbit was about 0.8-1.3 and 0.4-0.6 at $e_o$=0.02 and $e_o$=0.15, respectively. At calculations with a mass of planet c equal to a half of its present mass, such ratio was about 1.3-1.5 and 0.5-0.6. 
Based on the above ratios, it is possible to conclude that the total mass of planetesimals ejected into hyperbolic orbits could be about (3.5-7)$m_E$. The total mass of planetesimals in the feeding zone of planet c could exceed 10$m_E$ and 15$m_E$ at $e_o$=0.2 and $e_o$=0.15, respectively. 
Based on estimates of the amount of planetesimals ejected into hyperbolic orbits and on the integral of energy, Ipatov (2023b) concluded that the semi-major axis of the orbit of planet c could decrease by at least a factor of 1.5 during the growth of the mass of planet c from 3.5$m_E$ to 7$m_E$. 

\section{Delivery of planetesimals from the vicinity of the orbit of planet Proxima Centauri c to the inner planets  b and d}

The probability of a collision of a planetesimal initially located in the feeding zone of planet c with planet b was about 0.0002 and 0.001 at $e_o$ equal to 0.02 or 0.15, respectively (Ipatov, 2023b). 
The above values of the probabilities of collisions of planetesimals with planet b were greater than the probability of a collision with the Earth of a planetesimal migrated from the zone of the giant planets in the Solar System. The latter probability (per one planetesimal) was typically less than $10^{-5}$ (Marov, Ipatov, 2023). 
Planet d was not included in integrations, and the probability of collisions of planetesimals with this planet was calculated based on some formulas and on the arrays of orbital elements of migrated planetesimals during their evolution.
The amount of material delivered from the feeding zone of planet c to planet d could be about twice less than that delivered to planet b. A lot of icy material and volatiles could be delivered to planets b and d. 

\section{Motion of planetesimals in the outer part of the Hill sphere of the star Proxima Centauri}
About 90\% of the planetesimals that first reached 500 AU from the star, for the first time reached 1200 AU from the star in less than 1 million years (Ipatov, 2023c). The inclinations of orbits of 80\% of the planetesimals that moved between 500 or 1200 AU from the star did not exceed $10^o$. 
Mainly only the bodies that came into the Hill sphere of Proxima Centauri  from outside can move in strongly inclined orbits of bodies in the outer part of the Hill sphere of this star. The radius of the Hill sphere of the star Proxima Centauri is by an order of magnitude smaller than the radius of the outer boundary of the Hills cloud in the Solar System and is two orders of magnitude smaller than the radius of the Hill sphere of the Sun. Therefore, it is difficult to expect the existence of such a massive analogue of the Oort cloud near this star as near the Sun.

\section{Probabilities of collisions of planetesimals with planets in the TRAPPIST 1 exoplanetary system}
The exoplanetary system TRAPPIST-1 consists of a star with a mass equal to 0.0898 of the mass of the Sun and 7 planets (from b to h) with masses from 0.33 to 1.37 Earth's masses. The semi-major axes of the planets' orbits are in the range from 0.012 to 0.062 AU. 
The motion of planetesimals under the gravitational influence of the star and seven TRAPPIST-1 planets (from b to h) was calculated with the use of the symplectic code from (Levison, Duncan, 1994) similar to my studies of migration of planetesimals in the Proxima Centauri system.
In each considered variant of calculations, the semi-major axes of orbits of $N_o$ planetesimals were near the semi-major axis of one of the planets. Initial eccentricities of orbits of planetesimals were equal to 0.02 or 0.15. Their initial inclinations equaled to $e_o$/2 rad.  Some calculations with $N_o$=250 were discussed in (Ipatov, 2022). Below I present the results for $N_o$=1000 and a time integration step $t_s$=0.01 or $t_s$=0.02 days. 

Most of collisions of planetesimals with planets took place in less than 10 Kyr. For disks b-g more than a half of collisions were during the first 1 Kyr. However, last collisions of planetesimals with planets could be after a few million years. There were no collisions of planetesimals with the host star. Not more than 3\% of planetesimals were ejected into hyperbolic orbits. The fraction of planetesimals collided with the 'host' planet (compared to collisions with all planets) typically was smaller for a greater considered time interval. For disks c-h, the fraction $f_1$ of planetesimals that collided with the host planet was about 0.37-0.63  and  0.27-0.53 at $e_o$=0.02 and  $e_o$=0.15, respectively. 
The second and third values  of the fraction of planetesimals that collided with neighbouring planets are designated as $f_2$ and $f_3$.  The values of $f_2$ and $f_3$ 
were in the ranges 0.17-0.28 and 0.11-0.17 at $e_o$=0.02 and were 0.2-0.32 and 0.08-0.19 at $e_o$=0.15. For disk b and $e_o$=0.02, the value of $f_1$ was in the range 0.78-0.8, and $f_2$ and $f_3$ were between 0.18-0.2 and 0.01-0.012, respectively (the range is for calculations with an integration step $t_s$ equaled to 0.01 and 0.02). For disk b and $e_o$=0.15, the values of $f_1$,  $f_2$, and $f_3$ were 0.74,  0.22-0.23, and 0.015-0.021, respectively. The fraction of collisions of planetesimals with the 'host' planet was usually smaller for disks located farther from the star. In each variant, including the migration from a disk located beyond the outer disk h, collisions were with all planets. Therefore, the outer layers of neighboring planets in the TRAPPIST-1 system can include similar material if there were many planetesimals near their orbits at the late stages of planetary formation. 
Ipatov (1993, 2019) noted that  each terrestrial planet could incorporate planetesimals from the feeding zones of all terrestrial planets.

\section{Conclusions}
The motion of planetesimals was studied in the Proxima Centauri and TRAPPIST 1 exoplanetary systems. The size of the feeding zone of planet Proxima Centauri c is discussed. It was noted that after hundreds of Myrs, some planetesimals could still move in elliptical resonant orbits inside the feeding zone of this planet that had been mainly cleared from planetesimals. The probability of a collision of a planetesimal initially located in the feeding zone of planet c with inner planet b was obtained to be about 0.0002 and 0.001 at initial eccentricity of orbits of planetesimals equal to 0.02 or 0.15, respectively. A lot of icy material and volatiles could be delivered from the icy zone near the orbit of planet c to inner planets b and d. The inclinations of orbits of 80\% of the planetesimals that moved between 500 or 1200 AU from the star did not exceed $10^o$. It was obtained that several planets in the TRAPPIST-1 system accumulated planetesimals initially located at the same distance. Outer layers of neighbouring TRAPPIST-1 planets can include similar material. 
\section{Acknowledgements}
The studies were carried out under government-financed research project for the Vernadsky Institute of Geochemistry and Analytical Chemistry of Russian Academy of Sciences.

\end{document}